\documentclass[prb,aps,twocolumn,superscriptaddress,showpacs,footinbib,longbibliography]{revtex4-2}
\usepackage[colorlinks=true,linkcolor=black, citecolor=blue, urlcolor=blue, 
unicode=true]{hyperref}

\usepackage{graphicx}
\usepackage{epstopdf}
\usepackage{color}
\usepackage{amsmath}

\begin{document}

\title{Spin reorientations in structurally metastable, disordered, and hexagonal Cr$_{7}$Te$_{8}$}

\author{K. Guratinder*}
\affiliation{School of Physics and Astronomy, The University of Edinburgh, Edinburgh EH9 3JZ, United Kingdom}

\author{T.G. Romig*}
\affiliation{Department of Chemistry and  Biochemistry, University of Maryland, College Park, Maryland 20742, USA}

\author{H.C. Mandujano}
\affiliation{Department of Chemistry and  Biochemistry, University of Maryland, College Park, Maryland 20742, USA}

\author{C. Stock}
\affiliation{School of Physics and Astronomy,The University of Edinburgh, Edinburgh EH9 3JZ, United Kingdom}

\author{E. E. Rodriguez}
\affiliation{Department of Chemistry and  Biochemistry, University of Maryland, College Park, Maryland 20742, USA}
\affiliation{NIST Center for Neutron Research, 100 Bureau Drive, Gaithersburg, Maryland 20899, USA}

\date{\today}

\begin{abstract}

Vapor deposited two-dimensional Cr$_{7}$Te$_{8}$ displays unusual temperature dependent Hall effect properties, including a room temperature anomalous Hall effect, sign reversals of the Hall resistivity on cooling, and a peak in the Hall resistivity at low temperatures.  The two dimensional Cr$_{7}$Te$_{8}$ heterostructures that form the basis of these measurements are hexagonal in structure.  We study the magnetic and structural properties of bulk Cr$_{7}$Te$_{8}$ synthesized by quenching from 1000 $^{\circ}$C with the goal of relating the magnetic, structural, and electronic properties. This quenched phase is metastable, hexagonal, and displays different magnetic properties from the slow-cooled and more thermodynamically stable monoclinic phase. High-resolution x-ray diffraction of the quenched hexagonal phase finds a first-order transition to a lower symmetry monoclinic phase on \textit{heating} above $\sim$ 550 K.   Magnetic susceptibility measurements of the quenched hexagonal phase reveal ferromagnetic ordering above room temperature, along with the two distinct transitions at $\sim$ 220~K and $\sim$ 70~K.   Through neutron diffraction studies, we find the $\sim$ 220 K anomaly is a spin reorientation transition of the ferromagnetically aligned magnetic moments and the $\sim70$ K feature represents a transition from a high temperature ferromagnet to a low temperature antiferromagnet.  We suggest that these magnetic transitions are related to changes in the unit cell dimensions and are connected to the temperature dependent Hall resisitivity studied in two-dimensional heterostructures.   This implies a link between structural, magnetic, and electronic properties in the ``pseudo" two-dimensional chromium tellurides.

\end{abstract}

\pacs{}
\maketitle

\section{Introduction}

The study of two-dimensional (2D) van der Waals materials has attracted increasing interest due to their unique magnetic properties, with magnetic order persisting to the monolayer limit, and their potential applications in spintronics.~\cite{ma2024unusual, Yao24:18, Burch18:563, dijkstra1989band,gong2019two,han2018van,zhang2025advances} One recent advancement is in the construction of heterostructure devices by stacking multiple layers of 2D material candidates.~\cite{Zhong19:15}  These heterostructures enhance material functionality while reducing the physical footprint compared to the devices based on three-dimensional (3D) materials.\cite{sakthivel2023heterostructures, novoselov20162d} However, the limited number of known 2D materials, especially those that exhibit magnetic order at or above room temperature, has motivated researchers to explore hybrid heterostructures that effectively integrate both 2D and 3D materials.\cite{ma2024unusual, meng2024gate,younis2025magnetoresistance} One potential solution is to investigate materials that are pseudo-2D where there is a strong anisotropic behavior in two dimensions but no true van der Waals gap as in 2D systems \cite{miao2025tuning}.

In this regard, the chromium tellurides (Cr$_x$Te$_y$)~\cite{Ipser83:265} have garnered significant attention in both fundamental research and industry~\cite{Mondal25:17} due to their room-temperature ferromagnetism~\cite{Chua21:33,Zhang21:12}, retained to the monolayer limit~\cite{Sun21:11,Lee21:4}, and exceptional chemical stability under ambient conditions. In particular, the Hall effect in Cr$_{1+\delta}$Te$_{2}$, particularly in thin films, is sizeable and highly tuneable even changing sign on tuning the chemical parameter $\delta$.~\cite{Fujisawa23:35} These materials are intriguing due to the wide range of stoichiometric and symmetry variants accessible through slight modifications in synthetic conditions and sensitivity to Cr vacancies.  In compounds with reduced Cr content, such as Cr$_5$Te$_8$ and Cr$_7$Te$_8$, the Cr vacancies can be either randomly distributed or ordered, leading to different crystallographic symmetries—trigonal for random distributions and monoclinic for ordered arrangements.~\cite{Yang23:18}   The electronic properties of Cr$_x$Te$_y$ are equally diverse, spanning metallic, half-metallic, and semiconducting regimes~\cite{Zhang19:4}.  Vapor deposited two-dimensional Cr$_{7}$Te$_{8}$ displays unusual Hall\cite{ma2024unusual} properties, including a room temperature anomalous Hall effect, sign reversals of the Hall resistivity on cooling, and a peak in the Hall resistivity at low temperatures.  The low dimensional Cr$_{7}$Te$_{8}$ heterostructures that form the basis of these measurements are hexagonal in structure.~\cite{Yao24:18} This underscores a delicate interplay between crystal structure, Cr vacancy ordering, and electronic correlations.

The magnetic properties of Cr$_x$Te$_y$ are correspondingly varied and display a wide range of magnetic ground states--from ferromagnetic (FM) ordering with Curie temperatures ($T_\mathrm{C}$) between 150–340~K to canted antiferromagnetic (AFM) configurations—depending on composition, structure, and vacancy arrangement~\cite{ shimada1996photoemission, dijkstra1989band,huang2008neutron,andresen1970,ohsawa1972,haraldsen1937,lukoschus2004,goswami2024high}. Structurally, Cr$_x$Te$_y$  crystallize in a variety of space groups ranging from hexagonal Ni--As type structures with randomly distributed chromium vacancies to more ordered monoclinic and orthorhombic phases featuring systematic chromium ordering \cite{jiang2020magnetic, purwar2023anomalous}. Vacancy distribution not only stabilizes different crystal symmetries but also modulates interatomic distances and bonding characteristics, significantly impacting magnetic exchange interactions and electronic bandwidths \cite{dijkstra1989band}.  For example, Cr$_x$Te$_y$ compounds with a chromium-to-tellurium ($x\over y$) ratio greater than 0.75 exhibit ferromagnetism~\cite{cao2019structure,shimada1996photoemission}.   These observations highlight the role of vacancy ordering in determining the material’s properties and the need for full crystallographic and magnetic characterization.  

 The role of Cr site vacancy and ordering is highlighted by the specific example of  Cr$_7$Te$_8$ where the saturation magnetization is known to vary significantly depending on the vacancy configuration.  While magnetic properties of many  Cr$_x$Te$_y$ compounds, such as Cr$_2$Te$_3$, Cr$_5$Te$_8$, and Cr$_3$Te$_4$, are well characterized Cr$_7$Te$_8$  remains comparatively unexplored \cite{bian2021covalent, wang2024critical, goswami2024critical, ma2024unusual, mondal2019anisotropic}.  The magnetic structural properties of Cr$_{7}$Te$_{8}$ were first reported by Hashimoto \textit{et al.} in Ref. \onlinecite{hashimoto1969magnetic} and summarized here.    Crystallographically, Cr$_{7}$Te$_{8}$ can exist in two distinct structural forms based on its thermal treatment: an ordered and a disordered arrangement of vacancies among the successive chromium layers with the disordered phase being hexagonal and the ordered monoclinic.  These structural variations significantly impact the measured magnetic properties in  Ref. \onlinecite{hashimoto1969magnetic} with simple ferromagnetic behavior identified in the disordered hexagonal phase and an additional magnetic transition in the vacancy ordered monoclinic phase.   
 
The magnetic structure of Cr$_{7}$Te$_{8}$ remains unclear, particularly regarding the role of disordered vacancies on the magnetic, structural, electronic responses discussed above and in relation to the measured electronic Hall properties. This work addresses this relationship through x-ray and neutron diffraction and magnetic susceptibility measurements whereby we characterize the magnetic structure of polycrystalline hexagonal phase Cr$_{7}$Te$_{8}$.  The goal of this work is to establish a link between magnetic, structural, and the reported electronic properties. 
 
 \section{Experimental methods}
 
Polycrystalline hexagonal phase Cr$_{7}$Te$_{8}$ was prepared using high purity Cr (Thermo-Scientific 200 mesh 99.94$\%$) and Te (Thermo-Scientific 200 mesh 99.6$\%$) powders inside of an argon (Ar) filled glove box in a stoichiometric ratio of 7 parts Cr to 8 parts Te, the elements were ground in a mortar and pestle until finely mixed and transferred to a quartz ampoule. The solid mixture was then evacuated and sealed utilizing a hydrogen torch, where it was placed inside a box furnace and heated to 450$^{\circ}$C at a rate of 1 $^{\circ}$C/min and held for 12 hours. It was then heated to 1270$^{\circ}$C at a rate of 1$^{\circ}$ C/min and held for 1 hour; following this step it was cooled to 1000$^{\circ}$C where it was held for 60 hours before being quenched in an ice water bath.

Powder x-ray Diffraction (pXRD) was measured in reflection geometry utilizing a Bruker D8 instrument with Cu K-$\alpha$ wavelength and a LynxEye detector.   Further high resolution x-ray powder diffraction was performed using a Rigaku Smartlab equipment with a Cu source and a wavelength of $\lambda=$1.54 \AA\ fixed by a Johansson monochromator.  Measurements were done in reflection Bragg-Bretano geometry utilizing parabolic Cross Beam Optics (CBO).~\cite{Osakabe17:33}  Rietveld refinement of the pXRD pattern at room temperature confirmed the lattice parameters of the hexagonal unit cell with dimensions of \textit{a} = 4.0029(3)\r{A} and \textit{c} = 6.2392(2)\r{A}.   Corroborating the fractional chemical concentrations obtained from Rietveld refinement, elemental analysis was conducted with energy dispersive X-ray spectroscopy.

\begin{table}
	\caption{Refined structural parameters for hexagonal (quenched) Cr$_{7}$Te$_{8}$ at 300~K extracted from the Rietveld refinement of X-ray powder diffraction data on D8 Bruker. Space group, $P6_3/mmc$, No.~194, $a$ = 4.003(3)\AA, $c$ = 6.239(2)\AA, $R_p$ = 9.68 $\%$, and $wR_{p}$ =6.97$\%$.}
	\label{T1}
	\setlength{\tabcolsep}{2pt}
	\begin{tabular}{ccccccc}
		\hline
		Atom & $x$ & $y$ & $z$ & Occ. & U$_{iso}$ & Site \\ \hline
		Cr & 0.00  & 0.00  & 0.00  & 0.9(1) & 0.01(1) & 2$a$ \\ 
		Te & 1/3 & 2/3  & 1/4 & 1.00 & 0.001(1) & 2$c$ \\ 
		\hline
	\end{tabular}
\end{table}

 \begin{table} [b!]
	\caption{Refined structural parameters for monoclinic (slow-cooled) Cr$_{7}$Te$_{8}$ at 300~K extracted from the Rietveld refinement of X-ray powder diffraction data on D8 Bruker. Space group, $C2/m$, No.~12, $a$ = 14.009(3) \AA, $b$ = 3.940(1)\AA, $c$ = 6.877(1)\AA, $\beta$=118.2487(30)$^{\circ}$, $R_p$ = 9.308 $\%$, and $wR_p$ = 6.190$\%$.}
	\label{T2}
	\setlength{\tabcolsep}{2pt}
	\begin{tabular}{ccccccc}
		\hline
		Atom & $x$ & $y$ & $z$ & Occ. & U$_{iso}$ & Site \\ \hline
		Cr1 & 0.256(2)  & 0.500  & 0.274(2)  & 1.00 & 0.012(2) & 4$i$ \\ 
		Cr2 & 0.000  & 0.500  & 0.00  & 1.00 & 0.009(2) & 2$b$ \\
		Cr3 & 0.500  & 0.500  & 0.500  & 0.18(3) & 0.006(2) & 2$c$ \\
		Te1 & 0.368(1) & 0.500  & 0.031(2) & 1.00 & 0.010(3) & 4$i$ \\ 
		Te2 & 0.120(2) & 0.500  & 0.455(2) & 1.00 & 0.011(2) & 4$i$ \\
		\hline
	\end{tabular}
\end{table}

In terms of characterizing the magnetism, the macroscopic magnetic properties as a function of temperature were measured on a Quantum Design Magnetic Property Measurement System (MPMS).  To investigate the magnetic structure of Cr$_{7}$Te$_{8}$ we employed neutron powder diffraction using the BT-1 high resolution powder diffractometer (NIST, USA) with $\lambda$=2.0785 \AA. 

\section{Results and Discussion}

\subsection{Crystal Structure}

\begin{figure*}[t!]
	\centering\includegraphics[width=150mm,trim=0.0cm 0.2cm 0.0cm 0.5cm,clip=true]{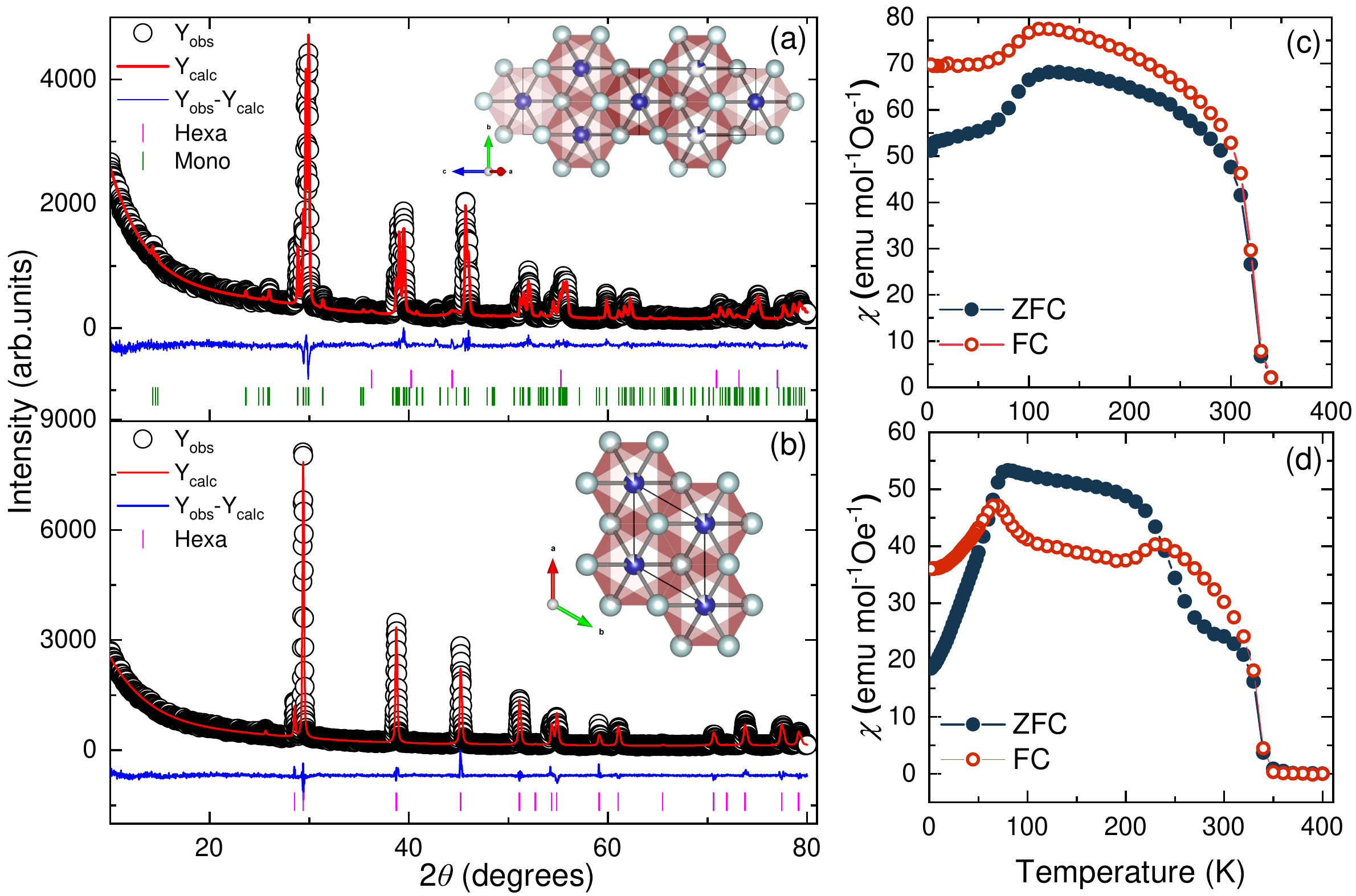}
	\caption{\label{fig_Diff1}
		Crystal Structure of Cr$_{7}$Te$_{8}$\ from the $c$-axis and crystal structure of Cr$_{7}$Te$_{8}$ with $(a)$ displaying the slow-cooled ordered monoclinic phase.    The hexagonal Ni-As type structure with disordered chromium vacancies from quenching is illustrated in $(b)$.   Results of the Rietveld refinement of Cr$_{7}$Te$_{8}$  are listed Tables 1 and 2.  The corresponding temperature dependent magnetic susceptibility for the ordered (slow-cooled) monoclinic phase $(c)$ and disordered (quenched) phase $(d)$ are plotted.}
\end{figure*}

As discussed previously, there are two different polymorphs of Cr$_{7}$Te$_{8}$. The disordered phase of Cr$_7$Te$_8$ crystallizes in a NiAs-type structure, featuring face-sharing octahedra around the tellurium anion. It adopts the high-symmetry space group $P6_3/mmc$ (No. 194), attributable to the presence of randomly distributed Cr vacancies~\cite{hashimoto1969magnetic,goswami2024critical}. In contrast, the ordered phase of this compound exhibits a lower-symmetry monoclinic structure, arising from systematic chromium vacancy ordering.  To stabilize the disordered vacancy phase, we employed a rapid quenching method, cooling the reaction vessel from 1000°C into an ice-water bath as detailed above. Through pXRD, we expect to be able identify this quenched disordered phase since it has a higher-symmetry over the ordered monoclinic phase.  This would result in a fewer number of Bragg reflections for the hexagonal quenched phase over the slow-cooled monoclinic (and ordered) phase which would have many more Bragg reflections owing to the lower symmetry.   This is demonstrated in Fig, \ref{fig_Diff1} $(a)$ and $(b)$ which present Rietveld refinements at room temperature of the monoclinic and hexagonal phase, respectively, as fit to the Bruker pXRD data at room temperature.  The structural parameters from this refinement are tabulated in Table \ref{T1} for the hexagonal (quenched) phase and compared to the slow cooled monoclinic refinements listed in Table \ref{T2}.  Their respective zero field cooled (ZFC) and field cooled (FC) magnetic susceptibility as a function of temperature plots are shown alongside Figs. \ref{fig_Diff1} $(c)$ and $(d)$ and are discussed  below.

\begin{figure*}
	\centering\includegraphics[width=150mm,trim=0.0cm 3.8cm 0.0cm 3.5cm,clip=true]{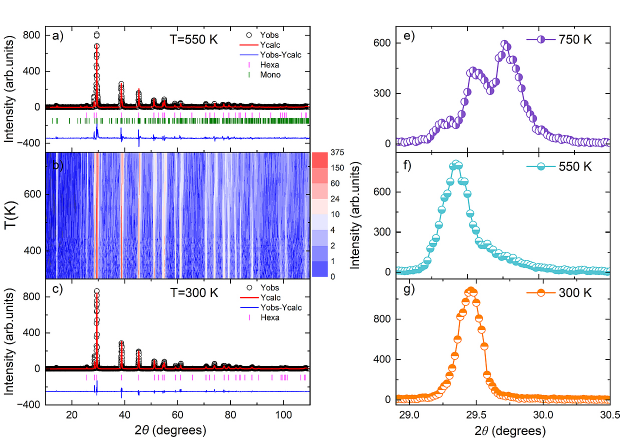}
	\caption{\label{fig_Diff2}
		Structure analysis using the Rietveld refinements of powder sample of Cr$_{7}$Te$_{8}$\ at two different temperatures using monochromatic Cu K-$\alpha$ radiation. $(a)$, $(c)$ demonstrate the phase change from hexagonal to mixture of monoclinic and hexagonal as a function of temperature and $(b)$ shows the peak splitting evident at higher angles supporting the phase change from hexagonal to mixed phase. The left panel $(e-f)$ shows the change in the peak as the phase changes from monoclinic to hexagonal as a function of temperature.  The coexistence of both phases at high temperatures is indicative of a first order transition from hexagonal to monoclinic on heating.} 
\end{figure*}

As measured with the Rigaku Smartlab equipped with the high-temperature furnace and monochromatic Cu K-$\alpha$ radiation, summarized in Fig. \ref{fig_Diff2}, the temperature-dependent pXRD patterns reveal the change of the hexagonal phase into its high-temperature thermodynamically stable monoclinic phase (Fig. \ref{fig_Diff2}) on warming. In addition to the contour plot of the multiple pXRD patterns (Fig. \ref{fig_Diff2} $b$), Fig. \ref{fig_Diff2} $(a,c)$ shows representative Rietveld refinements of the monoclinic and hexagonal phase respectively.  Peak broadening and splitting at higher angles as a function of temperature reveal the transition from high-symmetry to lower monoclinic symmetry phase on heating (Fig. \ref{fig_Diff2} $e-g$).  We observe both monoclinic and hexagonal phases at intermediate temperatures.  This coexistence implies a first-order transition from hexagonal to monoclinic on heating. The metastable hexagonal phase begins to undergo monoclinic distortion with increasing temperature and significant peak splitting is observed above T $\sim$ 550  K (Fig. \ref{fig_Diff2} $f$). The phase fraction of the monoclinic phase continues to grow up to the highest temperature of 750 K taken in our experiment.

Chromium tellurides are known to crystallize in various symmetries, ranging from the low-symmetry $C2/m$ to high-symmetry $P6_3/mmc$ phases, depending on stoichiometry and thermal history~\cite{bian2021covalent}. The observed peak splitting supports the hypothesis that the hexagonal phase undergoes a distortion pathway into the monoclinic structure, likely through subtle shifts in chromium occupancy or local strain fields. Notably, the crystallographic space group and the degree of Cr-vacancy ordering appear to exert a significant influence on the magnetic properties of the system. For instance, prior studies have linked chromium ordering to distinct magnetic transitions and changes in saturation magnetization~\cite{goswami2024critical, purwar2023anomalous}.  We now discusses the magnetic properties measured through bulk magnetic susceptibility in the context of the observed nuclear structure.

\subsection{Linking Magnetic Susceptibility with Structure}

\begin{figure}[h!]
	\centering\includegraphics[width=85mm,trim=0.0cm 0.1cm 0.0cm 0.1cm,clip=true]{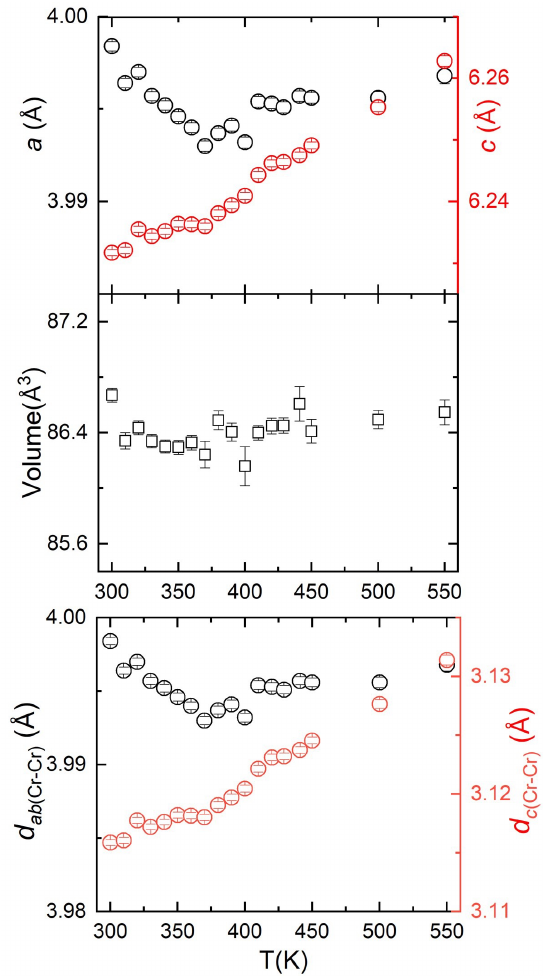}
    \caption{\label{Tdep_lattice}
		The temperature dependence of lattice constants, the volume and the Cr-Cr bond distance fitted against the hexagonal space group. A marked contraction is observed in the \textit{c} lattice and an expansion is measured in the \textit{a} lattice up to 360~K. }
		\end{figure}

The magnetic susceptibility comparing the hexagonal/disordered (quenched) and monoclinic/ordered (slow-cooled) phases of Cr$_{7}$Te$_{8}$ are illustrated in Figs. \ref{fig_Diff1} $(c,d)$.  Our results illustrate  the onset of high temperature ferromagnetic order in both polymorphs of Cr$_{7}$Te$_{8}$, however, and in contrast to early reports in Ref. \onlinecite{hashimoto1969magnetic}, further magnetic transitions are observed in the quenched (disordered) hexagonal phase with one at $\sim$ 220 K and a second lower temperature $\sim$ 70 K.  The upper $\sim$ 220 K retains a dominant ferromagnetic character while the lower transition of $\sim$ 70 K is indicative of a transition to antiferromagnetism, given the measured decrease in both field cooled (FC) and zero-field cooled (ZFC) susceptibilities.  This lower temperature antiferromagnetic transition observed in both polymorphs of Cr$_{7}$Te$_{8}$ is notably absent in lower-symmetry and lower-chromium-content compounds such as Cr$_5$Te$_8$ \cite{Wang2019}.

We investigate a possible link between structural and magnetic properties in Fig. \ref{Tdep_lattice} where we plot the hexagonal lattice constants as a function of temperature.  We observe an expansion along the $c$-axis at $\sim$ 375 K with the volume of the unit cell remaining constant within experimental error.   This expansion is close to the $\sim$ 350 K ferromagnetic ordering that we observe for hexagonal (quenched) Cr$_{7}$Te$_{8}$ (Fig. \ref{fig_Diff1} $d$).  This result is consistent with high-pressure neutron diffraction measurements on CrTe~\cite{Lambert78:39} and Cr$_{5}$Te$_{8}$~\cite{jiang2020magnetic} that find ferromagnetism disappears at high pressures.  This is further supported by first principles calculations in Refs. \onlinecite{dijkstra1989band,zhou2021strain,Kanomata98:177} that suggest that the itinerant magnetism observed in the chromium telluride system is heavily influenced by the Cr–Cr orbital overlap. 

We hypothesize that the complex behavior reflected in the magnetic susceptibility arises from thermal contraction on cooling in the \textit{ab}-plane, coupled with expansion along the \textit{c}-axis, which results in longer interlayer Cr–Cr distances.  This taken with high pressure measurements discussed above is suggestive that larger Cr-Cr overlap within the $a-b$ plane coupled with reduced overlap along $c$ is favorable for ferromagnetism in the Cr$_{x}$Te$_{y}$ series.  In this regard, the increased Curie temperature in Cr$_7$Te$_8$ may be attributed to the higher chromium content that enhances Cr–Cr orbital overlap and stabilizes the ferromagnetic state.   By extension, we further suggest that the unique lower temperature magnetic transitions observed at 70~K and 220 K may originate from the randomly distributed vacancies within the lattice. These vacancies can locally perturb the itinerant electron system, altering the magnetic exchange landscape and giving rise to possible new lower temperature magnetic ground states.  It is interesting to note that these magnetic transitions in disordered/hexagonal (quenched) Cr$_{7}$Te$_{8}$ correlate with the anomalous Hall effect, reinforcing the idea the connection between structural and electronic properties and this point is further discussed below~\cite{ma2024unusual}.  Having addressed the magnetization and its relation to structural properties, we now investigate the magnetic structure applying neutron diffraction

\subsection{Magnetic Structure from Neutron Powder Diffraction}

\begin{figure}[h!]
	\centering\includegraphics[width=75mm,trim=0.0cm 1.8cm 0.0cm 1.5cm,clip=true]
	{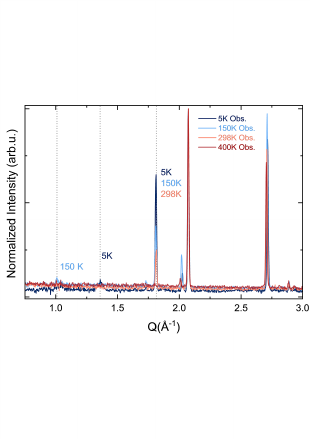}
	\caption{\label{fig_Combined}
		Combined neutron diffraction patterns of data taken at 5~K, 150~K, 298~K, and 400~K. A new anti-ferromagnetic contribution is observed in the 150~K diffraction data and a peak attributed to a \textit{k} vector of $(\frac{1}{2}, \frac{1}{2}, 0)$ in the 5~K data.}
\end{figure}

\begin{figure}[h!]
	\includegraphics[width=105mm,trim=1.0cm 0.0cm 0.0cm 0.0cm,clip=true]{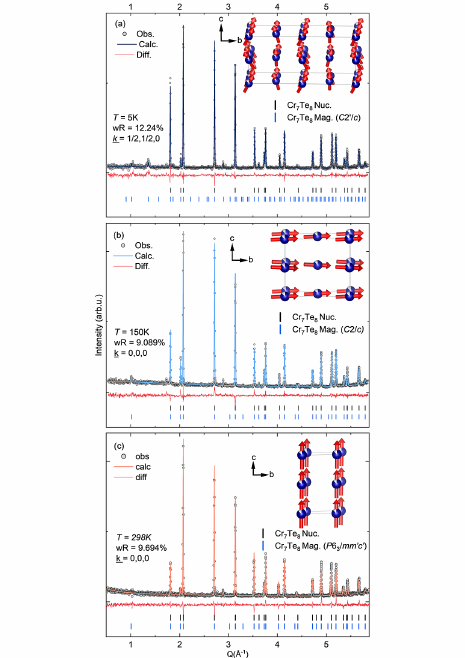}
	\caption{\label{fig_NPD}
		Neutron diffraction patterns at 5~K (a), 150~K (b), and 298~K (c) for a powder sample of Cr$_{7}$Te$_{8}$.The experimental data are drawn as grey dots. Rietveld refinement calculations are shown as Dark blue, light blue, and salmon for 5~K, 150~K, and 298~K respectively.}
\end{figure}

\begin{figure}[h!]
	\centering\includegraphics[width=75mm,trim=0.0cm 0.0cm 0.0cm 0.0cm,clip=true]
	{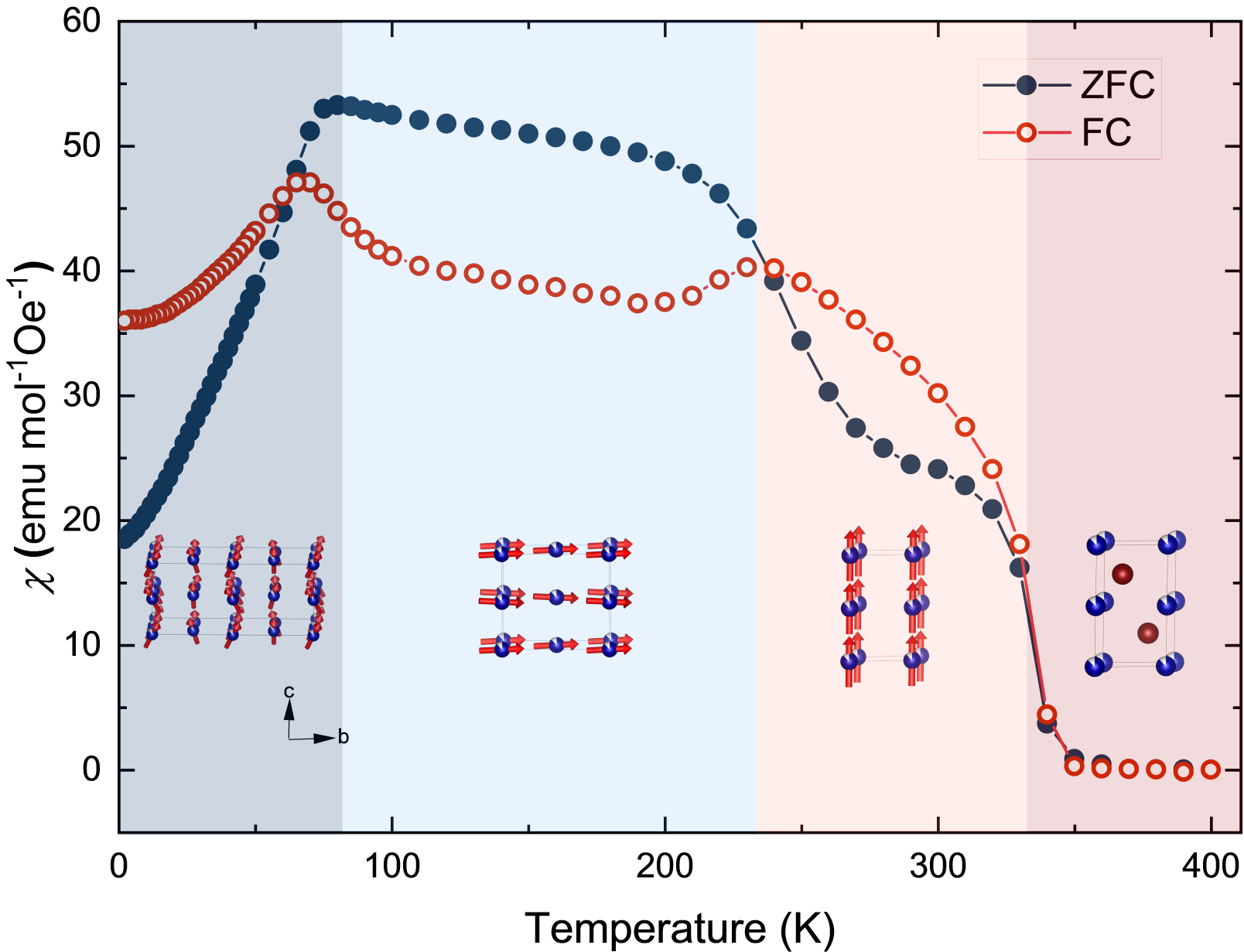}
	\caption{\label{fig_XvT}
		Zero field cooled and field cooled magnetic susceptibility measurements from 2~K-400~K. The ferromagnetic transition of $\sim$ 350~K can be seen as well as an additional magnetic transition at $\sim$ 220~K and finally a transition with anti-ferromagnetic contributions at $\sim$ 70~K. The magnetic structures displayed in the phase diagram are pulled from the refinement of the neutron diffraction data.} 
\end{figure}

To study the magnetic structure of Cr$_{7}$Te$_{8}$, neutron powder diffraction experiments were performed on the high-resolution neutron powder diffractometer BT-1 (NIST, USA) using a wavelength of $\lambda=2.0785$~\AA. The magnetization measurements for the quenched hexagonal phase in Fig. \ref{fig_Diff1} $(d)$ illustrate 4 temperature regimes guiding temperature-dependent diffraction measurements taken at 400~K in the paramagnetic phase and in the lower temperature magnetic phases at 298~K, 150~K, and 5~K.

Figure \ref{fig_Combined} presents the normalized diffraction patterns in these four different temperature regimes, illustrating magnetic moment formation evidenced by the emergence of new resolution limited Bragg peaks across varying temperatures.  We first summarize the results, starting with the low temperature data and then discuss in more detail the magnetic refinements below.    Ferromagnetic (characterized by a propagation wavevector of (0,0,0)) order would manifest as magnetic Bragg peaks overlapping with nuclear peaks. In contrast, antiferromagnetic order can lead to Bragg peaks at new positions in momentum, which are distinct from nuclear Bragg peaks. The 5 K data in Fig. \ref{fig_Combined} displays such new low-temperature peaks, which are forbidden by nuclear symmetry, thereby confirming the onset of antiferromagnetic (AFM)-like contributions at low temperatures.  In terms of momentum space, the additional peaks at T=5 K  are indexed by a \textit{k} vector of $(\frac{1}{2}, \frac{1}{2}, 0)$.   At 150~K, another new peak emerges, which can be explained by a reduction in magnetic symmetry to $C$2/$c$ with a \textit{k} vector of $(0, 0, 0)$.  At 298~K,  no new peaks appear that cannot be attributed to nuclear symmetry, indicating a simple ferromagnetic state.   The neutron diffraction data suggests three magnetic transitions:  a paramagnetic to a ferromagnetic state at $\sim$ 350~K; a transition to a canted ferromagnet with AFM-like contributions at 220~K in the \textit{ab} plane; and finally a transition at 70~K to a ferromagnet, again in the \textit{c}-axis with AFM-like contributions in the \textit{ab} plane. These transitions correspond to the structures resolved at 298~K 150~K and 5~K, respectively.  

The magnetic structure was solved with the BT-1 neutron powder diffraction (NPD) data by analyzing the possible magnetic symmetries provided by the $k$-SUBGROUPGMAG tool~\cite{Perez-Mato15:45} from the Bilbao Crystallographic Server within GSAS-II~\cite{Dreele24:80}.  All refinements were done utilizing the high symmetry $P$6$_3$/$mmc$ phase given the absence of observable monoclinic peaks in the diffraction pattern.   Figure \ref{fig_NPD} presents a summary of the refined magnetic diffraction patterns at three distinct temperatures 298~K, 150~K, and 5~K, characterizing the magnetic temperature regimes discussed above. The T=400~K data was utilized to define the nuclear diffraction pattern as no new magnetic peaks can be seen due to it being taken above the magnetic transition temperature and above the highest transition temperature measured with susceptibility discussed above.  The unit cell was solved to be \textit{a} = \textit{b} = 3.9998(2)\AA\ and \textit{c} = 6.2544(3)\AA\ with no magnetic contributions.  The refined magnetic moments from the low temperature magnetic phases are tabulated in Tables \ref{tab:ferro} and \ref{tab:antiferro}. The details of the magnetic structure are discussed in the appendix section.

\begin{table}[h!]
\caption{Magnetic moment of chromium at $298~\text{K}$ and $150~\text{K}$ solved from neutron diffraction.}
    \label{tab:ferro}
    \centering
    \begin{tabular}{ccccc}
        \hline
        $T$ & \textbf{$M_x$} & \textbf{$M_y$} & \textbf{$M_z$} & \textbf{$M_{tot}$} ($\mu_{B}$)\\ \hline
        298 K & 0.0         & 0.0         & 1.46(3) & 1.46(3)         \\ 
        150 K & 0.67(4)         & 2.84(3)         & 0.23(2) & 2.93(2)        \\ \hline
    \end{tabular}
\end{table}

\begin{table}[h!]
    \centering
    \caption{Magnetic moment of chromium at $5~\text{K}$ solved from neutron diffraction.  The average total magnetic moment across all sites is solved to be 2.94 $\mu_{B}$.}
    \label{tab:antiferro}
    \begin{tabular}{ccccc}
        \hline
        $Site$ & \textbf{$M_x$} & \textbf{$M_y$} & \textbf{$M_z$} & \textbf{$M_{tot}$} ($\mu_{B}$)\\  \hline
        Cr(1) & 1.4(3) & 0.05(1) & 2.5(1) & 2.8(2)  \\ 
        Cr(2) & 1.4(3) & 0.7(2) & 2.5(1) & 2.9(2) \\ 
        Cr(3) & 1.4(3) & 1.2(1) & 2.5(1) & 3.0(2) \\ 
        Cr(4) & 1.4(3) & 1.2(1) & 2.5(1) & 3.1(2) \\ \hline
    \end{tabular}
\end{table}

At room temperature, changes in the intensity of nuclear peaks allowed the structure to be resolved as a simple ferromagnet with the magnetic moment axis aligned along the \textit{c}-direction. Lattice parameter refinements revealed an expansion in the \textit{a}/\textit{b} parameters to 4.0034(9)~\AA\ and a contraction in the \textit{c} parameter to 6.24046(8)~\AA. This solved structure aligns with the susceptibility data, which demonstrates a ferromagnetic arrangement between $\sim$ 350~K and approximately 220~K that is out of plane. The room temperature measurements were best described by the magnetic space group $\textit{P}6_3/\textit{mm'c'}$, which exhibited the highest symmetry and achieved a weighted residual factor ($wR$) of 9.69\%. This fit is isostructural with the nuclear refinement, with a slight reduction in symmetry due to the alignment of the chromium magnetic moments along the \textit{c}-axis.

At 150~K, the appearance of a new diffraction peak near 1.01 \AA$^{-1}$, which is forbidden by the original symmetry group, indicates the emergence of an antiferromagnetic (AFM) component which refines to a monoclinic $C2/c$ magnetic space group. This observation aligns with the magnetic transition at approximately 220~K, as shown in the susceptibility data. The resolved structure at this temperature reveals a rotation of the ferromagnetic moment from pointing along the $c$-axis to the \textit{ab}-plane, accompanied by measurable AFM-like canting components along the \textit{c}-axis. Similar flips in magnetization easy axis have been reported in neutron studies of monoclinic Cr$_{5}$Te$_{8}$.\cite{huang2008neutron} We speculate that this behavior to the thermodynamic contraction of the \textit{c} lattice parameter at lower temperatures.

Upon cooling to 5~K, a new diffraction peak at 1.34 \AA$^{-1}$ emerges, which cannot be accounted for with a \textit{k} vector of $(0, 0, 0)$. Fits using \textit{k} vectors of $(0, 0, \frac{1}{3})$ and $(\frac{1}{2}, \frac{1}{2}, 0)$ were explored. It was determined that the new magnetic peak arises from a symmetry reduction to $C$2'/$c$ from the $C$2/$c$ magnetic space group at 150~K, with a propagation vector of $(\frac{1}{2}, \frac{1}{2}, 0)$. At this temperature, the magnetic moments undergo a further easy-axis transition, reorienting along the \textit{c}-axis. Additionally, enhanced AFM contributions are observed within the \textit{ab}-plane, influencing the chromium moments.

 We finally summarize these results and the magnetic structure refinements in comparison to the magnetic susceptibility discussed above in Fig. \ref{fig_XvT}.  Through the combination of magnetic susceptibility and neutron diffraction we find magnetic transitions at 220~K and 70~K tied to reorientation of the spins, as well as the transition to the ferromagnetic state at $\sim$ 350~K.  Finally, the 400~K data in Fig. \ref{fig_Combined} shows only the nuclear structure, as the crystal system lies above the Curie temperature and exhibits no magnetic contributions to the diffraction pattern.  Our magnetic refinements from neutron diffraction connect a change of the magnetic structure with each observable anomaly in the magnetic susceptibility. 

\section{Discussion}

We have presented a magnetic and structural study of the metastable hexagonal phase of Cr$_{7}$Te$_{8}$. High-resolution X-ray diffraction measurements clearly demonstrate temperature-driven peak splitting, indicative of a progressive symmetry reduction from the metastable hexagonal $P$6$_3$/$mmc$ phase towards a lower symmetry monoclinic $C$2/$m$ phase on warming to higher temperatures. This monoclinic phase gradually dominates at elevated temperatures, with notable peak splitting occurring beyond $\sim$ 550~K, suggesting a thermally induced structural transformation.  We emphasize that this transition is unusual given that typically on lowering temperature a symmetry breaking occurs with a reduction in the number of symmetry elements in the structural space group.  In the case investigated here, the opposite occurs with symmetry being lowered on heating.  We note that the hexagonal phase does not reappear on cooling indicative of its metastability and that the monoclinic phase is more thermodynamically stable.

Magnetic characterization through neutron powder diffraction across a broad temperature range (400~K to 5~K) elucidates the magnetic phase transitions inherent to Cr$_{7}$Te$_{8}$.  The transition from a paramagnetic state above $\sim$ 350~K to a ferromagnetic state below this Curie temperature is unambiguously confirmed by susceptibility data and neutron refinements, which identify a simple ferromagnetic ordering with moments aligned primarily along the crystallographic $c$-axis at room temperature. Cooling below 150~K induces a magnetic transition characterized by a reorientation of the magnetic moment direction and the emergence of canting in the magnetic moments, signifying an onset of competing magnetic interactions. At low temperatures, neutron diffraction experiments at  5~K observe additional antiferromagnetic-like canting described by a magnetic propagation vector $\left(\frac{1}{2}, \frac{1}{2}, 0\right)$.  

The two magnetic transitions observed in susceptibility measurements correspond well with these neutron diffraction findings.  We further observe on cooling an expansion of the $c$ unit cell length and a contraction within the $ab$ plane, which we suggest drives the magnetic ordering in Cr$_{7}$Te$_{8}$.  Given the absence of these transitions in the structurally ordered material, we suggest that the presence of randomly distributed vacancies and chromium site disorder likely perturbs exchange interactions, driving the lower temperature complex magnetic phases and transitions observed. 

Having connected the structural and magnetic properties of Cr$_{7}$Te$_{8}$, we now discuss the connection with the electronics.  As noted at the beginning of this paper, an anomaly is observed in the Hall conductivity for vapor transport grown ultrathin sheets of Cr$_{7}$Te$_{8}$.~\cite{ma2024unusual}   In particular a clear anomaly is observed near the onset of antiferromagnetic ordering at $\sim$ 70 K.  Give the connection between Berry phase and the spin orientation with the magnetism breaking time-reversal-symmetry,  this is highly suggestive of a link between the static magnetic structure and the electronic response probed through the Hall effect.  Given the proposed connection between the lattice strain (lattice constants) and the magnetic structure, strain would be expected to impact electronic transport as indeed suggested in Cr$_{2}$Te$_{3}$ based on first principles calculations and microscopy~\cite{Chi23:14}.   

\section{Conclusion}

This study establishes a connection between the structural, magnetic, and electronic properties of metastable, hexagonal Cr$_{7}$Te$_{8}$. Through magnetic susceptibility and neutron diffraction, we identified several low-temperature magnetic transitions and linked them to spin reorientation transitions. A particularly notable finding is the transition to an antiferromagnetic component at low temperatures, which we propose is related to anomalies observed in the Hall resistivity.  The chromium telluride (Cr$_{x}$Te$_{y}$) system provides a unique platform for studying the coupling between magnetic and nuclear structural properties with electronic responses.

\section*{Acknowledgments}

We gratefully acknowledge valuable assistance from R. Rae with the Rigaku smartlab facility for the X-ray measurements in CSEC at the University of Edinburgh, UK. We are indebted to A. Ali (MPI-FKF) for insightful discussions and the feedback. EER, TR, and HCM thank the U.S. Department of Energy (DOE), Office of Science (Grant No. DE-SC0016434), for the financial support.

\bibliographystyle{apsrev4-2}
%

\section*{Data Availability}
The data that support the findings of this study are available from the corresponding author upon request. ~\cite{GuratinderCr7Te8_2025}.
\appendix
\section{{Additional Magnetic and Symmetry Analysis for Cr$_7$Te$_8$}}

\subsection{{Irreducible representations and basis vectors}}
Representation analysis was performed for the experimentally determined
crystal structures using standard symmetry methods. In both cases, the
propagation vector is $\mathbf{k}=(0,\,0,\,0)$. For the Cr site, the
magnetic representation decomposes into irreducible representations
that differ between 300~K and 150~K. The selected irreducible
representations correspond to moments constrained parallel to the
$c$ axis at 300~K and symmetry-allowed canted components within the
basal plane with a dominant component along the monoclinic $b$ axis at
150~K. These basis sets were employed to construct and refine the
magnetic configurations reported in the main text.

For the hexagonal phase of Cr$_7$Te$_8$, crystallizing in space group
\textit{P}6$_3$/$mmc$, the relevant irreducible representations of the
little group at $\mathbf{k}=(0,\,0,\,0)$ were obtained using standard
representation analysis and used as the starting point for the magnetic
refinements as mentioned in the main text.

\subsection*{300 K}
At 300~K, two Cr atoms are present in the crystallographic unit cell.
However, these atoms belong to the same Wyckoff position and are related
by symmetry operations of the space group \textit{P}6$_3$/$mmc$. As a
result, they form a single crystallographic orbit and therefore share a
single magnetic representation.

The magnetic representation for the Cr site decomposes as:

\begin{equation}
\Gamma_{\mathrm{mag}} = \Gamma_{3} + \Gamma_{7} + \Gamma_{9} + \Gamma_{11}.
\end{equation}

Magnetic refinements based on neutron diffraction data show that the
experimental results are best described by the irreducible
representation $\Gamma_{3}$, which constrains the magnetic moments to
lie parallel to the crystallographic $c$-axis.

\begin{table}[h]
\caption{(300~K) Basis vectors of the irreducible
representation $\Gamma_{3}$ for the Cr($2a$) site of Cr$_7$Te$_8$ in space
group $P$6$_3$/$mmc$ with propagation vector
$\mathbf{k}=(0,\,0,\,0)$.}
\label{tab:bv_300K_gamma3}

\begin{ruledtabular}
\begin{tabular}{ccc}
Atom &  Position&Basis vector $\psi$ \\
\hline
Cr1 &  $\begin{pmatrix} 0 \\ 0 \\ 0  \end{pmatrix}$&$\begin{pmatrix} 0 \\ 0 \\ 1 \end{pmatrix}$ \\
Cr2 &  $\begin{pmatrix} 0 \\ 0 \\ 1/2 \end{pmatrix}$&$\begin{pmatrix} 0 \\ 0 \\ 1 \end{pmatrix}$ \\
\end{tabular}
\end{ruledtabular}
\end{table}

\subsection*{150 K}

At 150~K, symmetry lowering results in multiple Cr atomic positions in
space group \textit{C}2/$c$, while the propagation vector remains
$\mathbf{k}=(0,\,0,\,0)$. Transforming from hexagonal to monoclinic space group changes the unit cell parameters to  $a$ = 4.01 \AA, $b$ = 6.94 \AA, $c$ = 6.22 \AA, and $\beta$ = 90$^{\circ}$. This transformation was performed using the ISOTROPY suit built within GSAS-II. 

In this case, there is a single Cr site with four atomic positions (two within the primitive unit cell). For this case, the decomposition of the magnetic representation is given by:
\begin{equation}
\Gamma_{\mathrm{mag}} = 3\Gamma_{1} + 3\Gamma_{3}.
\end{equation}

The ferromagnetic favorable irreducible representation ($\Gamma_{1}$) was used to describe this magnetic structure.

\begin{table}
\caption{(150~K) Basis vectors $\psi_{n}$ of the irreducible representation $\Gamma_{1}$ of
Cr$_7$Te$_8$ in space group \textit{C}2/$c$ with propagation vector
$\mathbf{k}=(0,\,0,\,0)$. }
\label{tab:bv_150K_site12_gamma1}
\begin{ruledtabular}
\begin{tabular}{ccccc}
Atom  &  Position   &  $\psi_{1}$    &  $\psi_{2}$    &  $\psi_{3}$   \\ \hline \\ 
Cr1  &  $\begin{pmatrix} 0 \\ 0 \\ 0 \end{pmatrix}$ &  $\begin{pmatrix}  1  \\  0  \\  0  \end{pmatrix}$&  $\begin{pmatrix}  0  \\  1  \\  0  \end{pmatrix}$&  $\begin{pmatrix}  0  \\  0  \\  1  \end{pmatrix}$\\  
Cr2  &  $\begin{pmatrix} 0 \\ 0 \\ 1/2 \end{pmatrix}$ &  $\begin{pmatrix} -1  \\  0  \\  0  \end{pmatrix}$&  $\begin{pmatrix}  0  \\  1  \\  0  \end{pmatrix}$&  $\begin{pmatrix}  0  \\  0  \\ -1  \end{pmatrix}$\\ 
\end{tabular}
\end{ruledtabular}
\end{table}

\subsection*{5 K}
When describing the 5 K magnetic structure phase, we find that not all Cr atoms correspond to independent crystallographic orbits once the monoclinic unit cell is quadrupled due to the magnetic propagation vector $\mathbf{k}=(1/2,\,1/2,\,0)$. Several atomic positions are related by symmetry and therefore share the same magnetic representation. Consequently, the magnetic representation analysis yields four equivalent decompositions corresponding to four inequivalent Cr sites.

The magnetic representation decomposes as follows:

\textbf{Sites 1 - 4:}
\begin{equation}
\Gamma_{\mathrm{mag}} = 3\Gamma_{1} + 3\Gamma_{3}.
\end{equation}

In the basis-vector tables below, Cr sites correspond to symmetry-related atomic positions belonging to the same crystallographic site and therefore share an identical magnetic basis vector.

\begin{table}
\caption{(5~K) Basis vectors $\psi_{n}$ of the irreducible representation $\Gamma_{3}$ for Cr in space group \textit{C}2/$c$. The four crystallographic sites (Cr1--Cr4) are listed sequentially.}
\label{bv_table}
\begin{ruledtabular}
\begin{tabular}{ccccc}
Atom & Position & $\psi_{1}$ & $\psi_{2}$ & $\psi_{3}$ \\ \hline

\multicolumn{5}{c}{\textbf{Site Cr1}} \\ \hline 

Cr1 &
$\begin{pmatrix} 1/4 \\ 1/4 \\ 0 \end{pmatrix}$ &
$\begin{pmatrix} 1 \\ 0 \\ 0 \end{pmatrix}$ &
$\begin{pmatrix} 0 \\ 1 \\ 0 \end{pmatrix}$ &
$\begin{pmatrix} 0 \\ 0 \\ 1 \end{pmatrix}$ \\

Cr2 &
$\begin{pmatrix} 3/4 \\ 1/4 \\ 1/2 \end{pmatrix}$ &
$\begin{pmatrix} 1 \\ 0 \\ 0 \end{pmatrix}$ &
$\begin{pmatrix} 0 \\ -1 \\ 0 \end{pmatrix}$ &
$\begin{pmatrix} 0 \\ 0 \\ 1 \end{pmatrix}$ \\ \hline

\multicolumn{5}{c}{\textbf{Site 2}} \\ \hline

Cr3 &
$\begin{pmatrix} 1/4 \\ 1/4 \\ 1/2 \end{pmatrix}$ &
$\begin{pmatrix} 1 \\ 0 \\ 0 \end{pmatrix}$ &
$\begin{pmatrix} 0 \\ 1 \\ 0 \end{pmatrix}$ &
$\begin{pmatrix} 0 \\ 0 \\ 1 \end{pmatrix}$ \\

Cr4 &
$\begin{pmatrix} 3/4 \\ 1/4 \\ 0 \end{pmatrix}$ &
$\begin{pmatrix} 1 \\ 0 \\ 0 \end{pmatrix}$ &
$\begin{pmatrix} 0 \\ -1 \\ 0 \end{pmatrix}$ &
$\begin{pmatrix} 0 \\ 0 \\ 1 \end{pmatrix}$ \\ \hline

\multicolumn{5}{c}{\textbf{Site 3}} \\ \hline 
Cr5 &
$\begin{pmatrix} 1/2 \\ 1/2 \\ 0 \end{pmatrix}$ &
$\begin{pmatrix} 1 \\ 0 \\ 0 \end{pmatrix}$ &
$\begin{pmatrix} 0 \\ 1 \\ 0 \end{pmatrix}$ &
$\begin{pmatrix} 0 \\ 0 \\ 1 \end{pmatrix}$ \\

Cr6 &
$\begin{pmatrix} 1/2 \\ 1/2 \\ 1/2 \end{pmatrix}$ &
$\begin{pmatrix} 1 \\ 0 \\ 0 \end{pmatrix}$ &
$\begin{pmatrix} 0 \\ -1 \\ 0 \end{pmatrix}$ &
$\begin{pmatrix} 0 \\ 0 \\ 1 \end{pmatrix}$ \\ \hline

\multicolumn{5}{c}{\textbf{Site 4}} \\ \hline 
Cr7 &
$\begin{pmatrix} 1/2 \\ 0 \\ 0 \end{pmatrix}$ &
$\begin{pmatrix} 1 \\ 0 \\ 0 \end{pmatrix}$ &
$\begin{pmatrix} 0 \\ 1 \\ 0 \end{pmatrix}$ &
$\begin{pmatrix} 0 \\ 0 \\ 1 \end{pmatrix}$ \\

Cr8 &
$\begin{pmatrix} 1/2 \\ 0 \\ 1/2 \end{pmatrix}$ &
$\begin{pmatrix} 1 \\ 0 \\ 0 \end{pmatrix}$ &
$\begin{pmatrix} 0 \\ -1 \\ 0 \end{pmatrix}$ &
$\begin{pmatrix} 0 \\ 0 \\ 1 \end{pmatrix}$ \\

\end{tabular}
\end{ruledtabular}
\end{table}


\begin{table*}[h]
\caption{Irreducible representations of Cr$_{7}$Te$_{8}$ for space group \textit{P6$_3$/mmc} and propagation vector $\mathbf{k}=(0,0,0)$ at 300 K.}
\small
\label{irrep_table}
\begin{tabular*}{\textwidth}{@{\extracolsep{\fill}} ccccccccccccccccccccccccc}
\hline\hline
      &  $1$    &  $3^+_z$    &  $3^-_z$    &  $2_{1,z}$    &  $6^-_{3,z}$    &  $6^+_{3,z}$    &  $2_{xx0}$    &  $2_x$    &  $2_y$    &  $2_{x\bar{x}0}$    &  $2_{x2x0}$    &  $2_{2xx0}$    &  $\bar{1}$    &  $\bar{3}^+_z$    &  $\bar{3}^-_z$    &  $m_z$    &  $\bar{6}^-_z$    &  $\bar{6}^+_z$    &  $m_{xx0}$    &  $m_x$    &  $m_y$    &  $c_{x\bar{x}0}$    &  $c_{x2x0}$    &  $c_{2xx0}$   \\ \hline 
$\Gamma_{1}$ &  1  &  1  &  1  &  1  &  1  &  1  &  1  &  1  &  1  &  1  &  1  &  1  &  1  &  1  &  1  &  1  &  1  &  1  &  1  &  1  &  1  &  1  &  1  &  1  \\
$\Gamma_{2}$ &  1  &  1  &  1  &  1  &  1  &  1  &  1  &  1  &  1  &  1  &  1  &  1  & -1  & -1  & -1  & -1  & -1  & -1  & -1  & -1  & -1  & -1  & -1  & -1  \\
$\Gamma_{3}$ &  1  &  1  &  1  &  1  &  1  &  1  & -1  & -1  & -1  & -1  & -1  & -1  &  1  &  1  &  1  &  1  &  1  &  1  & -1  & -1  & -1  & -1  & -1  & -1  \\
$\Gamma_{4}$ &  1  &  1  &  1  &  1  &  1  &  1  & -1  & -1  & -1  & -1  & -1  & -1  & -1  & -1  & -1  & -1  & -1  & -1  &  1  &  1  &  1  &  1  &  1  &  1  \\
$\Gamma_{5}$ &  1  &  1  &  1  & -1  & -1  & -1  &  1  &  1  &  1  & -1  & -1  & -1  &  1  &  1  &  1  & -1  & -1  & -1  &  1  &  1  &  1  & -1  & -1  & -1  \\
$\Gamma_{6}$ &  1  &  1  &  1  & -1  & -1  & -1  &  1  &  1  &  1  & -1  & -1  & -1  & -1  & -1  & -1  &  1  &  1  &  1  & -1  & -1  & -1  &  1  &  1  &  1  \\
$\Gamma_{7}$ &  1  &  1  &  1  & -1  & -1  & -1  & -1  & -1  & -1  &  1  &  1  &  1  &  1  &  1  &  1  & -1  & -1  & -1  & -1  & -1  & -1  &  1  &  1  &  1  \\
$\Gamma_{8}$ &  1  &  1  &  1  & -1  & -1  & -1  & -1  & -1  & -1  &  1  &  1  &  1  & -1  & -1  & -1  &  1  &  1  &  1  &  1  &  1  &  1  & -1  & -1  & -1  \\
$\Gamma_{9}$ &
$\begin{pmatrix}  1 & 0 \\ 0 & 1 \end{pmatrix}$ &
$\begin{pmatrix} -0.500 + 0.866i & 0 \\ 0 & -0.500 - 0.866i \end{pmatrix}$ &
$\begin{pmatrix} -0.500 - 0.866i & 0 \\ 0 & -0.500 + 0.866i \end{pmatrix}$ &
$\begin{pmatrix}  1 & 0 \\ 0 & 1 \end{pmatrix}$ &
$\begin{pmatrix} -0.500 + 0.866i & 0 \\ 0 & -0.500 - 0.866i \end{pmatrix}$ &
$\begin{pmatrix} -0.500 - 0.866i & 0 \\ 0 & -0.500 + 0.866i \end{pmatrix}$ &
$\begin{pmatrix}  0 & 1 \\ 1 & 0 \end{pmatrix}$ &
$\begin{pmatrix}  0 & -0.500 - 0.866i \\ -0.500 + 0.866i & 0 \end{pmatrix}$ &
$\begin{pmatrix}  0 & -0.500 + 0.866i \\ -0.500 - 0.866i & 0 \end{pmatrix}$ &
$\begin{pmatrix}  0 & 1 \\ 1 & 0 \end{pmatrix}$ &
$\begin{pmatrix}  0 & -0.500 - 0.866i \\ -0.500 + 0.866i & 0 \end{pmatrix}$ &
$\begin{pmatrix}  0 & -0.500 + 0.866i \\ -0.500 - 0.866i & 0 \end{pmatrix}$ &
$\begin{pmatrix}  1 & 0 \\ 0 & 1 \end{pmatrix}$ &
$\begin{pmatrix} -0.500 + 0.866i & 0 \\ 0 & -0.500 - 0.866i \end{pmatrix}$ &
$\begin{pmatrix} -0.500 - 0.866i & 0 \\ 0 & -0.500 + 0.866i \end{pmatrix}$ &
$\begin{pmatrix}  1 & 0 \\ 0 & 1 \end{pmatrix}$ &
$\begin{pmatrix} -0.500 + 0.866i & 0 \\ 0 & -0.500 - 0.866i \end{pmatrix}$ &
$\begin{pmatrix} -0.500 - 0.866i & 0 \\ 0 & -0.500 + 0.866i \end{pmatrix}$ &
$\begin{pmatrix}  0 & 1 \\ 1 & 0 \end{pmatrix}$ &
$\begin{pmatrix}  0 & -0.500 - 0.866i \\ -0.500 + 0.866i & 0 \end{pmatrix}$ &
$\begin{pmatrix}  0 & -0.500 + 0.866i \\ -0.500 - 0.866i & 0 \end{pmatrix}$ &
$\begin{pmatrix}  0 & 1 \\ 1 & 0 \end{pmatrix}$ &
$\begin{pmatrix}  0 & -0.500 - 0.866i \\ -0.500 + 0.866i & 0 \end{pmatrix}$ &
$\begin{pmatrix}  0 & -0.500 + 0.866i \\ -0.500 - 0.866i & 0 \end{pmatrix}$ \\
$\Gamma_{10}$ &
$\begin{pmatrix}  1 & 0 \\ 0 & 1 \end{pmatrix}$ &
$\begin{pmatrix} -0.500 + 0.866i & 0 \\ 0 & -0.500 - 0.866i \end{pmatrix}$ &
$\begin{pmatrix} -0.500 - 0.866i & 0 \\ 0 & -0.500 + 0.866i \end{pmatrix}$ &
$\begin{pmatrix}  1 & 0 \\ 0 & 1 \end{pmatrix}$ &
$\begin{pmatrix} -0.500 + 0.866i & 0 \\ 0 & -0.500 - 0.866i \end{pmatrix}$ &
$\begin{pmatrix} -0.500 - 0.866i & 0 \\ 0 & -0.500 + 0.866i \end{pmatrix}$ &
$\begin{pmatrix}  0 & 1 \\ 1 & 0 \end{pmatrix}$ &
$\begin{pmatrix}  0 & -0.500 - 0.866i \\ -0.500 + 0.866i & 0 \end{pmatrix}$ &
$\begin{pmatrix}  0 & -0.500 + 0.866i \\ -0.500 - 0.866i & 0 \end{pmatrix}$ &
$\begin{pmatrix}  0 & 1 \\ 1 & 0 \end{pmatrix}$ &
$\begin{pmatrix}  0 & -0.500 - 0.866i \\ -0.500 + 0.866i & 0 \end{pmatrix}$ &
$\begin{pmatrix}  0 & -0.500 + 0.866i \\ -0.500 - 0.866i & 0 \end{pmatrix}$ &
$\begin{pmatrix} -1 & 0 \\ 0 & -1 \end{pmatrix}$ &
$\begin{pmatrix} 0.500 - 0.866i & 0 \\ 0 & 0.500 + 0.866i \end{pmatrix}$ &
$\begin{pmatrix} 0.500 + 0.866i & 0 \\ 0 & 0.500 - 0.866i \end{pmatrix}$ &
$\begin{pmatrix} -1 & 0 \\ 0 & -1 \end{pmatrix}$ &
$\begin{pmatrix} 0.500 - 0.866i & 0 \\ 0 & 0.500 + 0.866i \end{pmatrix}$ &
$\begin{pmatrix} 0.500 + 0.866i & 0 \\ 0 & 0.500 - 0.866i \end{pmatrix}$ &
$\begin{pmatrix}  0 & -1 \\ -1 & 0 \end{pmatrix}$ &
$\begin{pmatrix}  0 & 0.500 + 0.866i \\ 0.500 - 0.866i & 0 \end{pmatrix}$ &
$\begin{pmatrix}  0 & 0.500 - 0.866i \\ 0.500 + 0.866i & 0 \end{pmatrix}$ &
$\begin{pmatrix}  0 & -1 \\ -1 & 0 \end{pmatrix}$ &
$\begin{pmatrix}  0 & 0.500 + 0.866i \\ 0.500 - 0.866i & 0 \end{pmatrix}$ &
$\begin{pmatrix}  0 & 0.500 - 0.866i \\ 0.500 + 0.866i & 0 \end{pmatrix}$ \\
$\Gamma_{11}$ &
$\begin{pmatrix}  1 & 0 \\ 0 & 1 \end{pmatrix}$ &
$\begin{pmatrix} -0.500 + 0.866i & 0 \\ 0 & -0.500 - 0.866i \end{pmatrix}$ &
$\begin{pmatrix} -0.500 - 0.866i & 0 \\ 0 & -0.500 + 0.866i \end{pmatrix}$ &
$\begin{pmatrix} -1 & 0 \\ 0 & -1 \end{pmatrix}$ &
$\begin{pmatrix} 0.500 - 0.866i & 0 \\ 0 & 0.500 + 0.866i \end{pmatrix}$ &
$\begin{pmatrix} 0.500 + 0.866i & 0 \\ 0 & 0.500 - 0.866i \end{pmatrix}$ &
$\begin{pmatrix}  0 & 1 \\ 1 & 0 \end{pmatrix}$ &
$\begin{pmatrix}  0 & -0.500 - 0.866i \\ -0.500 + 0.866i & 0 \end{pmatrix}$ &
$\begin{pmatrix}  0 & -0.500 + 0.866i \\ -0.500 - 0.866i & 0 \end{pmatrix}$ &
$\begin{pmatrix}  0 & -1 \\ -1 & 0 \end{pmatrix}$ &
$\begin{pmatrix}  0 & 0.500 + 0.866i \\ 0.500 - 0.866i & 0 \end{pmatrix}$ &
$\begin{pmatrix}  0 & 0.500 - 0.866i \\ 0.500 + 0.866i & 0 \end{pmatrix}$ &
$\begin{pmatrix}  1 & 0 \\ 0 & 1 \end{pmatrix}$ &
$\begin{pmatrix} -0.500 + 0.866i & 0 \\ 0 & -0.500 - 0.866i \end{pmatrix}$ &
$\begin{pmatrix} -0.500 - 0.866i & 0 \\ 0 & -0.500 + 0.866i \end{pmatrix}$ &
$\begin{pmatrix} -1 & 0 \\ 0 & -1 \end{pmatrix}$ &
$\begin{pmatrix} 0.500 - 0.866i & 0 \\ 0 & 0.500 + 0.866i \end{pmatrix}$ &
$\begin{pmatrix} 0.500 + 0.866i & 0 \\ 0 & 0.500 - 0.866i \end{pmatrix}$ &
$\begin{pmatrix}  0 & 1 \\ 1 & 0 \end{pmatrix}$ &
$\begin{pmatrix}  0 & -0.500 - 0.866i \\ -0.500 + 0.866i & 0 \end{pmatrix}$ &
$\begin{pmatrix}  0 & -0.500 + 0.866i \\ -0.500 - 0.866i & 0 \end{pmatrix}$ &
$\begin{pmatrix}  0 & -1 \\ -1 & 0 \end{pmatrix}$ &
$\begin{pmatrix}  0 & 0.500 + 0.866i \\ 0.500 - 0.866i & 0 \end{pmatrix}$ &
$\begin{pmatrix}  0 & 0.500 - 0.866i \\ 0.500 + 0.866i & 0 \end{pmatrix}$ \\
$\Gamma_{12}$ &
$\begin{pmatrix}  1 & 0 \\ 0 & 1 \end{pmatrix}$ &
$\begin{pmatrix} -0.500 + 0.866i & 0 \\ 0 & -0.500 - 0.866i \end{pmatrix}$ &
$\begin{pmatrix} -0.500 - 0.866i & 0 \\ 0 & -0.500 + 0.866i \end{pmatrix}$ &
$\begin{pmatrix} -1 & 0 \\ 0 & -1 \end{pmatrix}$ &
$\begin{pmatrix} 0.500 - 0.866i & 0 \\ 0 & 0.500 + 0.866i \end{pmatrix}$ &
$\begin{pmatrix} 0.500 + 0.866i & 0 \\ 0 & 0.500 - 0.866i \end{pmatrix}$ &
$\begin{pmatrix}  0 & 1 \\ 1 & 0 \end{pmatrix}$ &
$\begin{pmatrix}  0 & -0.500 - 0.866i \\ -0.500 + 0.866i & 0 \end{pmatrix}$ &
$\begin{pmatrix}  0 & -0.500 + 0.866i \\ -0.500 - 0.866i & 0 \end{pmatrix}$ &
$\begin{pmatrix}  0 & -1 \\ -1 & 0 \end{pmatrix}$ &
$\begin{pmatrix}  0 & 0.500 + 0.866i \\ 0.500 - 0.866i & 0 \end{pmatrix}$ &
$\begin{pmatrix}  0 & 0.500 - 0.866i \\ 0.500 + 0.866i & 0 \end{pmatrix}$ &
$\begin{pmatrix} -1 & 0 \\ 0 & -1 \end{pmatrix}$ &
$\begin{pmatrix} 0.500 - 0.866i & 0 \\ 0 & 0.500 + 0.866i \end{pmatrix}$ &
$\begin{pmatrix} 0.500 + 0.866i & 0 \\ 0 & 0.500 - 0.866i \end{pmatrix}$ &
$\begin{pmatrix}  1 & 0 \\ 0 & 1 \end{pmatrix}$ &
$\begin{pmatrix} -0.500 + 0.866i & 0 \\ 0 & -0.500 - 0.866i \end{pmatrix}$ &
$\begin{pmatrix} -0.500 - 0.866i & 0 \\ 0 & -0.500 + 0.866i \end{pmatrix}$ &
$\begin{pmatrix}  0 & -1 \\ -1 & 0 \end{pmatrix}$ &
$\begin{pmatrix}  0 & 0.500 + 0.866i \\ 0.500 - 0.866i & 0 \end{pmatrix}$ &
$\begin{pmatrix}  0 & 0.500 - 0.866i \\ 0.500 + 0.866i & 0 \end{pmatrix}$ &
$\begin{pmatrix}  0 & 1 \\ 1 & 0 \end{pmatrix}$ &
$\begin{pmatrix}  0 & -0.500 - 0.866i \\ -0.500 + 0.866i & 0 \end{pmatrix}$ &
$\begin{pmatrix}  0 & -0.500 + 0.866i \\ -0.500 - 0.866i & 0 \end{pmatrix}$ \\
\hline\hline
\end{tabular*}

\end{table*}

\begin{table}[h]
\caption{Irreducible representations of Cr for space group \textit{C}2/$c$ and propagation vector $\mathbf{k}$ = (0, 0, 0) at 150~K.}
\label{irrep_table_150K}
\begin{ruledtabular}
\begin{tabular}{c|cccc}
      &  $1$    &  $2_y$    &  $\bar{1}$    &  $c_y$   \\ \hline \\ 
$\Gamma_{1}$ & $ 1 $  & $ 1 $  & $ 1 $  & $ 1 $ \\ \\ 
$\Gamma_{2}$ & $ 1 $  & $ 1 $  & $-1 $  & $-1 $ \\ \\ 
$\Gamma_{3}$ & $ 1 $  & $-1 $  & $ 1 $  & $-1 $ \\ \\ 
$\Gamma_{4}$ & $ 1 $  & $-1 $  & $-1 $  & $ 1 $ \\ 
\end{tabular}
\end{ruledtabular}
\end{table}

\begin{table}[h]
\caption{Irreducible representations of Cr for space group $C$2/$c$ and propagation vector $\mathbf{k}$ = (0, 0, 0) at 5 K after duplicating $a$ and $b$ axes from $\mathbf{k}$ = (1/2, 1/2, 0).}
\label{irrep_table}
\begin{ruledtabular}
\begin{tabular}{c|cccc}
      &  $1$    &  $2_y$    &  $\bar{1}$    &  $c_y$   \\ \hline \\ 
$\Gamma_{1}$ & $ 1 $  & $ 1 $  & $ 1 $  & $ 1 $ \\ \\ 
$\Gamma_{2}$ & $ 1 $  & $ 1 $  & $-1 $  & $-1 $ \\ \\ 
$\Gamma_{3}$ & $ 1 $  & $-1 $  & $ 1 $  & $-1 $ \\ \\ 
$\Gamma_{4}$ & $ 1 $  & $-1 $  & $-1 $  & $ 1 $ \\ 
\end{tabular}
\end{ruledtabular}
\end{table}

\end{document}